# Design and Implementation of an Antenna Model for the Cooja simulator

Vishwesh Rege
Robert Bosch Centre for Cyber Physical Systems
Indian Institute of Science (IISc)
Bangalore, India
vishwesh.rege@cps.iisc.ernet.in

*Abstract*—**COOJA is a network simulator developed for wireless sensor networks. It can be used for high-level algorithm development as well as low-level device driver implementations for accurate simulation of wireless sensor networks before deployment. However, in a simulation Cooja assumes that the nodes are only equipped with omnidirectional antennas. There is currently no support for directional antennas. Due to the growing interest in the use of directional or smart antennas in wireless sensor networks, a model that can support directional antennas is essential for the realistic simulations of protocols relying on directional communication. This paper presents work on extending COOJA with a directional antenna model.**

## I. INTRODUCTION

Omnidirectional antennas radiate power uniformly in all directions. In other words, they have a 360° beamwidth or circular radiation pattern. Directional antennas on the other hand focus their radiation towards a particular direction. This usually provides a greater range at the cost of reduced beamwidth. Apart from the increased range while transmission, directional antennas have an added advantage of rejecting unwanted interferences, from interferers not in the coverage area of the antenna. As a result, they have increased signal to interference and noise ratio (SINR), which results in a lower bit error rate (BER) and also an improvement in capacity. Using directional antennas, neighboring nodes may be able to communicate simultaneously depending on the direction of transmission. This can increase the spatial reuse of the channel.

To exploit the characteristics of directional antennas, MAC and network layer protocols have to be designed by taking into account these physical layer effects. Several MAC [4, 5], localization [6, 7], and routing [8, 9], protocols using directional antennas have been proposed in the literature for ad hoc and sensor networks. The capacity and end-to-end delay of such networks have been shown to improve considerably by designing protocols and applications that exploit the characteristics of directional antennas. A simulator model that supports directional communication will greatly simplify the future development and evaluation of such protocols.

Cooja [10] is a sensor network simulator that can simulate the operation of networked embedded operating systems such as Contiki [12] and TinyOS [13]. In this work, the Cooja simulator is modified to implement antennas which can transmit in a specific direction. The modifications that have been made to support such directional communication are discussed. The developed model is flexible and can be easily modified to simulate any desired antenna pattern.

## II. SIMULATION MODEL

This section introduces the relevant components of the Cooja simulator and the modifications that have been made to incorporate support for directional communication.

### A. Direction interface

COOJA *interfaces* can be interfaces to hardware of simulated node devices (E.g. Button, LEDs) or interfaces to node properties (E.g. node position). Interfaces allow the user to analyze and interact with the simulated nodes. For example, the user can use the button interface to press a button and trigger an interrupt on the node similar to pressing the button in real life. Interfaces such as "Position" and "ID" on the other hand, only serve to provide the user more details about the node state. The user can observe the values of the position interface by moving the node around in the Network window.

Similar to the "Position" interface, an extra interface "Direction" is added, which represents directional properties of the simulated node's antenna. The "Direction" interface encapsulates the antenna properties such as type, orientation angle, gain and beamwidth.

***Type*** indicates the selected node's antenna radiation pattern type: Omni/Directional.

***Orientation*** indicates the direction in which the antenna's main lobe is pointing (in degrees; valid only for directional radiation pattern).

***Gain*** indicates the gain of the antenna and is calculated for any direction based on the angle with respect to the Orientation (maximum in the direction of the main lobe).

***Beamwidth*** indicates the angular width of the main lobe (in degrees).

All these values can be changed manually by the user any time during the simulation.

*B. Realistic radio medium*

Other components of the simulator can register as observers of *interfaces* and be notified when the properties of the interface change. A *radio medium* in Cooja is registered as an observer of each node's "Radio" interface. This allows the radio medium to be notified whenever a node's radio starts transmitting. The radio medium then decides based on each radio's location, which of the radios available in the simulation should receive the transmitted message.

The "Unit Disk Graph Medium (UDGM): Distance Loss" radio medium in Cooja was modified and used to implement and test the new directional property of nodes. The existing UDGM radio medium uses two distance parameters:

1. TRANSMITTING_RANGE to decide whether two nodes are within range, and

2. INTERFERENCE_RANGE to decide whether the transmission of one node can interfere with the reception of the other.

A node can receive a packet from a sender only if it is within its radius, defined by TRANSMITTING_RANGE. If the node's distance from the sender is greater than TRANSMITTING_RANGE but less than INTERFERENCE_RANGE, it can't receive the sender's packet but will be interfered by the sender's transmission. Outside the interference range, a node is oblivious to the sender's transmissions. Inside a node's transmitting radius, the receiver's signal strength is related to distance from the sender only by a distance factor defined as the ratio of the distance to the TRANSMITTING_RANGE.

As can be seen, the existing UDGM model doesn't accurately describe the path loss that is observed in real life, which is given in its simplest terms by the Friis transmission equation [3] (all quantities are in dB):

$$P_{rx} = P_{tx} + G_{tx} + G_{rx} - PL \qquad (1)$$

where,

$P_{rx}$ is the received signal power at the receive antenna,
$P_{tx}$ is the power delivered to the transmit antenna,
$G_{tx}$ is the gain of the transmitting antenna,
$G_{rx}$ is the gain of the receving antenna, and
PL is the free-space path loss

The UDGM radio medium was modified to realistically model the received signal strength using equation (1) and then determine whether the packet can be received based on the receiver sensitivity (which can be set by the user). The receiver sensitivity is defined as the minimum signal power required to successfully receive the packet. Thus, the receiver can receive the packet only if the signal strength at the receiver, $P_{rx}$ is greater than the receiver sensitivity. Another sender can interfere with a receiver's reception only if the signal strength of the interference is above a certain threshold.

## III. CALCULATION OF GAIN

The particular value of transmitter and receiver gains to be used in equation (1) depend on each of the antenna's radiation pattern, the orientation of the antenna and also the coordinates of the two nodes. The communication angles between the transmitter and receiver are calculated using the transmitter and receiver orientations and the relative coordinates of the two nodes. This angle is used to calculate the respective antenna gain based on the antenna specific radiation pattern.

The angle that the receiver makes with the transmitter is calculated as atan2 (x, y) where (x, y) are the relative coordinates between the two nodes. The gain in the receiver's direction is then calculated by considering this angle with respect to the transmitter antenna's orientation. A similar procedure is followed for calculating the receiver gain.

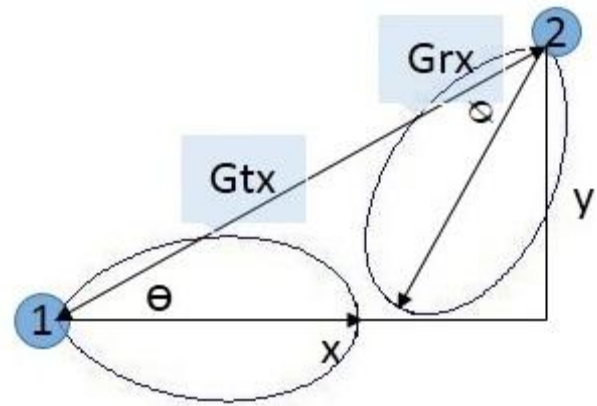

**Figure 1: Calculation of Gain**

Figure 1 illustrates the gain calculation. Node 1 is oriented at angle of 0° from the horizontal axis and node 2 is oriented at angle of -120° (say). So, the gain of node 1 in the direction of node 2 is calculated with reference to 0°, and the gain of node 2 in the direction of node 1 is calculated with reference to -120°.

Therefore,

ɵ = atan2 (x, y)

∝ = atan2 (x, y) − (−120)

(ɵ and ∝ are the angles with respect to the direction of maximum gain)

Note that in Figure 1, the point at which the line joining the two nodes intersects the antenna radiation pattern indicates the gain of the antenna for that transmission. Also note that the observed radiation pattern is only indicative of the gain and doesn't represent the coverage area of the node's transmission.

The gain value at a particular direction depends on the antenna radiation pattern. The model allows the radiation pattern to be changed easily as desired by only changing the gain equation. Figures 2-3 demonstrate a couple of the common radiation patterns as can be implemented in Cooja.

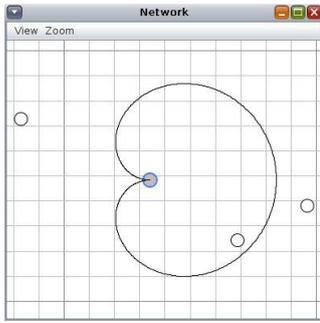

Figure 2: Cardioid radiation pattern

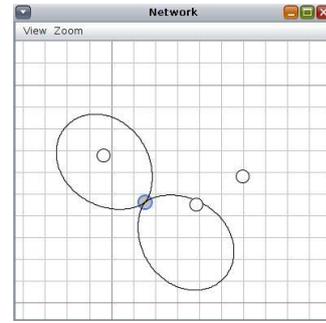

Figure 3: Dipole radiation pattern

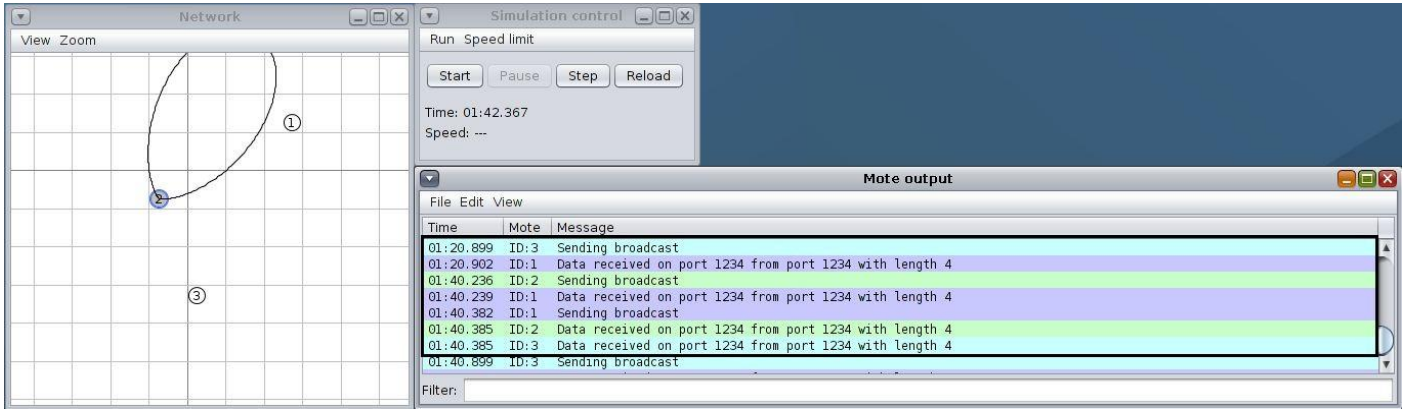

Figure 4: Cooja Simulation

## IV. MODEL EVALUATION

The model implementation was evaluated by comparing simulation results with expected results. 3 nodes were used in the simulation as shown in Figure 4 – two with omnidirectional antennas and one with a directional antenna. All of the nodes are programmed to broadcast packets at random intervals. Node 2 (with the directional antenna) has its orientation set to 60º with reference to the horizontal. Due to the nature of node placement (as seen in the figure), there should be no communication link between node 2 and node 3. The following observations are made:

Node 1's packets are successfully received by both node 2 and 3.
Node 2's packets are received by node 1 but not by node 3.
Node 3's packets are received by node 1 but not by node 2.

As can be seen, the observations agree with expected results, considering the directional properties of node 2.

Next, node 3 is moved down a few meters such that it is out range of node 1. Even when node 1 and 3 transmit simultaneously in this case, node 1's transmission is successfully received at node 2. This proves that the interference by node 3 at node 2 is effectively blocked out.

## V. CONCLUSION

An extension to the Cooja simulator is proposed for modeling the effect of directional antennas. The proposed model allows selection of any radiation pattern for representing the transmit/receive gain of a node in the simulation. Additionally, modifications are made to the radio model to accurately represent the link budget of a wireless link, according to Friis equation. Finally, the model is evaluated by analyzing the communication of nodes in the simulated network.

For future work, it would be useful to link the antenna model with the node's hardware interface such as to allow control of the antenna through code running on the simulated nodes. This will permit the simulation of protocols that make use of adaptive gain control methods.